Controlling phonon squeezing and correlation via one- and two-phonon interference

Oleg V. Misochko[1*], Jianbo Hu[2,3], Kazutaka G. Nakamura[2,3]

[1]*Institute of Solid State Physics, Russian Academy of Sciences*

*142432 Chernogolovka, Moscow region, Russia*

[2]*Materials and Structures Laboratory, Tokyo Institute of Technology*

*R3-10, 4259 Nagatsuta, Yokohama 226-8503, Japan*

[3]*JST-CREST, Sanban-cho bldg, 5, Sanban-cho, Chiyoda, Tokyo 102-0075, Japan*

*Corresponding author: e-mail: misochko@issp.ac.ru;  phone: ++7(495) 9628054;  fax:    ++7(496) 5249701

**Abstract**

When ultrafast laser pulse strikes the crystal with a van Hove singularity in the phonon density of states, it can create a pair of anti-correlated in wave-vector phonons. As a result, the atomic fluctuations in either position or momentum become squeezed in such a way that their size might fall below the vacuum level. The ultrafast pulses can also generate a biphonon state in which the constituent phonons are correlated and/or entangled. Here we show that via the interplay between one- and two-phonon interference the bound and squeezed two-phonon state in (110) oriented ZnTe single crystal can be manipulated.

1.     **Introduction**

Recent advances in laser technology have led to the generation of laser pulses whose duration is shorter than the time of many motions in solids. The use of such ultrashort pulses has made possible direct, time-resolved observation of elementary microscopic motions such as vibrations of crystal lattice (phonons). To date the coherent phonons have been observed in a great variety of solids [1-5] with increasing attention being turned towards not only observation but also understanding of physics which occurs at this short time scale. In analogy with photon states in quantum optics [6-8] the generation and manipulation of specific phonon states has become a major issue in condensed matter physics [9-20]. Yet, notwithstanding the recent experimental and theoretical advances, a clear relation between quantum mechanical coherent phonon state and the experimentally observed coherent phonons, typically described in classical terms, is still missing. In attempting to fill the gap, there has been a growing interest in the study of fluctuation properties of the coherent phonons. The fluctuations of one-phonon coherent excitations are masked by the large coherent amplitude that behaves classically and thus can be reduced to zero. Therefore, the generation of squeezed phonons by either second-order Raman scattering [9,10,12] or a pair of phase-locked pulses [11] has been theoretically predicted and signatures of phonon squeezing have been observed in a number of experiments [12-18]. It should be noted that some doubt was raised as to whether the vacuum squeezing for phonons was actually achieved in these experiments [10, 19, 20].

Owing to a well-defined phase, the phonons generated by an ultrafast pulse can be coherently controlled [5, 11]. Although interference, due to which such a control is achieved, is intrinsically a classical phenomenon, the superposition principle which underlies it is also at the heart of quantum mechanics. Indeed, in some interference experiments we encounter the idea of quantum entanglement, which became clear after the famous paper by Einstein, Podolsky and Rosen [21] had singled out some startling features of the quantum mechanics. Schrödinger emphasized that these features are due to the existence of what he called "entangled states," which are two-particle states that cannot be factored into products of two single-particle states in any representation [22]. "Entanglement" is just Schrödinger's name for superposition in a composite system. A typical example of the composite excitation in the case of crystal lattice is a two-phonon bound (biphonon) state [23] observed recently in the time-domain for zinc telluride [24]. Therefore, in this work we have attempted to study the two-phonon bound and squeezed states created by ultrafast laser pulses in ZnTe and observed that the degree of squeezing and correlation of these composite lattice excitations can be manipulated.

## 2. Experimental

The sample used in this study is a single crystal of zinc telluride with (110) surface. ZnTe has a cubic crystal structure of the zinc-blende type $T_d(\bar{4}\bar{3}m)$ with two atoms per unit cell. Its first-order Raman spectrum consists of two lines: The highest in frequency (6.3 THz) is denoted the longitudinal optical (LO) branch and the lowest (5.3 THz) is the doubly degenerate transverse optical (TO) branch. For our crystal orientation, the first-order Raman scattering in ZnTe is weak for TO($\Gamma$)- and prohibited for LO($\Gamma$)-phonons due to the resonance and symmetry conditions, respectively [24]. The second order Raman scattering involves two phonons. Because of this the conservation law does not require that the scattering take place at the zone center. However, the density of two phonon states tends to be greater for larger phonon wave vectors and thus most second order scattering take place at the Brillouin zone boundary. Critical points for the ZnTe reciprocal lattice occur at $X(1,0,0)$, $L(\frac{1}{2},\frac{1}{2},\frac{1}{2})$, and $W(1,\frac{1}{2},0)$, while the ZnTe band gap at room temperature is 2.25eV.

Our experimental set-up is shown in Fig.1. The method used is ultrafast pump-probe technique in which the strong pump pulse drives the crystal into an excited time-varying state, which perturbs the weaker probe pulse that follows behind. The detected signal is the transmitted intensity of the probe beam as a function of the time delay $t$ of the probe relative to the pump pulse. In brief, we excite the crystal creating phonons in a time short compared to phonon lifetimes and their inverse frequencies. The first condition means that we are dealing with transient lattice state, while the second is responsible for coherent nature of the lattice excitation. Given that our difference-detection scheme is in some respects similar to homodyne detection, the balancing of the detectors eliminates the contribution of noise that is equally distributed between the quadratures and thus allows measuring only the fluctuations induced by laser pulse. Admittedly, to create lattice coherence one has to establish fixed phase relations either among different $q$ modes of the same phonon branch [25] or between vacuum and excited states of a single $q = 0$ phonon mode [26]. In transparent materials, this can be achieved exclusively owing to the large spectral bandwidth of ultrashort laser pulses that generate nonstationary phonon states by mixing different lattice states through impulsive stimulated Raman process. For the impulsive excitation by a single ultrafast pulse such Raman-like models [9,10,12] predict the generation of squeezed phonons whenever the second order Raman tensor is nonvanishing, see, however [11].

To excite and detect the coherent phonons we used a mode-locked Ti-sapphire laser generating pulses of 40 fs duration centered at 800.0 nm (1.55 eV). Since the photon energy was lower than the ZnTe band gap, the excitation had off-resonant character. The laser had a repetition rate of 86 MHz and provided average power of a 55 mW for the pump and a 2 mW for the probe pulse both focused to a 50-µm diameter spot. The polarizations of the pump and probe beam were orthogonal. For coherent control experiments, the stronger pump pulse was fed into a modified Mach-Zehnder interferometer to be divided into two pulses with a variable interpulse separation.

## 3. Results and discussion

Following the impulsive optical excitation, coherent part of the transient transmission in (110) ZnTe crystal at room temperature starts to oscillate [24] as shown in Fig. 2(a). To understand the oscillatory nature of the transient transmission let us recall that before the pump pulse strikes the crystal, the lattice atoms are moving with random phases relative to one another. Ultrafast excitation delivers an impulsive force whose magnitude is proportional to the displacement of a given atom from its equilibrium in such a way that atoms with a larger displacement receive a larger kick toward the equilibrium [12, 13]. The directional change towards the equilibrium is due to two-phonon excitation with the net result bringing the atomic fluctuations more in step with one another. Right after the pump pulse, the amplitudes of displacement fluctuations are reduced at the expense of stretched fluctuations in momentum as schematically shown in Fig.3.

The change in transmission is induced by second-order Raman scattering [9, 10, 12] $\Delta T \equiv T - T_0 = \sum_q \frac{\partial^2 T}{\partial Q_q \partial Q_{q'}} \langle Q_q Q_{q'} \rangle$ , where $Q$ is normal mode operator and the average is over the phonon states. Because of phase coherent excitation the average is reduced to $\langle Q_q Q_{-q} \rangle = \langle \Delta u^2(\pm q,t) \rangle$ with $u$ being the atomic displacement. Consequently, what is observed by the probe pulse is shaped by, and dependent on the fluctuations of the centre mass of the lattice unit cell, that is the width of displacement distribution, see Fig.3.

### 3.1. Single pump pulse excitation

The oscillation lifetime and frequency in the (110) oriented ZnTe after single pump pulse excitation are 1.4 ps and 3.67 THz, respectively, the latter being close to, but a bit higher than that of 2TA(X) overtone in thermally excited ZnTe crystal [24], see Fig.2(b,c). Moreover, the oscillatory signal being independent of pump polarization [24] is consistent with an entity having the full $\Gamma_1$ symmetry: one of the representations appropriate for two transverse acoustic phonons from X-point of the Brillouin zone - $\Gamma_1 \otimes \Gamma_{12} \otimes \Gamma_{15}$. A positive frequency shift of $\cong 0.5$ THz, as compared to the 2TA(X) frequency, exposes a bound state of the two-phonon mode produced by a repulsive interaction [23, 24, 27, 28]. It is the divergence in the density of states at the zone boundary due to van Hove singularity that allows effectively pump energy into acoustic band resulting in the creation of a bound state, and it is residual phonon-phonon interaction that causes the bound state to split off from above the band of free two-phonon states.

We see that after the interaction with ultrafast laser pulse the crystal lattice in ZnTe is excited into two acoustic phonon modes $|q_+\rangle$ and $|q_-\rangle$ characterized by equal frequencies and equal but opposite wave vectors. By virtue of the conservation principles that govern their creation, the "twin" $|q_+\rangle$ and $|q_-\rangle$ phonons interfere destructively resulting in two-phonon coherence. Each phonon mode is in a superposition of its vacuum and first excited state [26], whereas the sum of their coherent amplitudes is reduced to zero due to phase coherent, simultaneous excitation [12]. Physically, this means that the $|q_+\rangle$ and $|q_-\rangle$ phonons can be more correlated with each other than with themselves: $2|\langle \hat{a}_+ \hat{a}_- \rangle| > \langle \hat{a}_+^\dagger \hat{a}_+ \rangle + \langle \hat{a}_-^\dagger \hat{a}_- \rangle$, where $\hat{a}_\pm$ and $\hat{a}_\pm^\dagger$ are the annihilation and creation operators for $|q_+\rangle$ and $|q_-\rangle$ phonons, respectively. Moreover, due to their simultaneous creation, the quantum states of these two-phonon modes are such that each one of the "twin" phonons cannot be described without referring to the other one. Since transverse acoustic phonons are doubly degenerate the two-phonon state is $|\varphi\rangle_\pm = \frac{1}{\sqrt{2}}(|V_+\rangle|V_-\rangle \pm |H_+\rangle|H_-\rangle)$, where $|V_i\rangle$ and $|H_i\rangle$ refer to the state in which $|q_i\rangle$ phonon is, respectively, vertically and horizontally polarized. If now a measurement is performed at $|q_+\rangle$, according to the outcome, the $|q_+\rangle$ will collapse to one of the eigenstates $|V_+\rangle$ or $|H_+\rangle$. But, and this is the essential point, immediately also the $|q_-\rangle$ will collapse to $|V_-\rangle|$ or $|H_-\rangle$, respectively. This means that a measurement on one constituent phonon can immediately influence another phonon. Thus, after a measurement on phonon one, we can instantaneously predict the result of a future measurement performed on phonon two.

We see that the constituent phonons, apart from being anti-correlated in wave vector, are entangled in polarization. Notice that in this case the lattice excitation in ZnTe is similar to two qubits. Such a superposition of two macroscopically different $|q_+\rangle$ and

$|q_-\rangle$ states often referred to as ''Schrödinger' cats" may exhibit both squeezing and entanglement [11, 29]. In addition, the correlated or entangled nature of the composite lattice excitation can be affected by the binding of its constituent phonons. Indeed, the biphonon state, reminiscent of a Cooper pair in superconductor (or more exactly of the Shafroth paired electrons [30]) cannot be reduced to two *independent* phonons as its appearance is accompanied by a gapped low energy spectrum after the phonon localization [31]. It should be noted that the entanglement is a special type of correlation that can be shared only among quantum objects, and, furthermore, coherence and entanglement are almost mutually exclusive properties: Coherence relies on separability, while entanglement excludes it.

Alternatively, we can consider the coherent lattice oscillations in ZnTe as a superposition between vacuum and first excited state of the crystal after "phonon localization" [24, 28]. Remember, the quantity that is localized at the formation of biphonon is the distance between the two phonons since the density-density correlation function is localized but not the wave function itself. Here, making use of the ideas developed by a number of authors [9,10,12], the coherent lattice excitation created in ZnTe by ultrafast laser pulse can be described as a two-phonon squeezed state defined by $|\alpha_+,\alpha_-,\zeta\rangle = \hat{D}_+(\alpha_+)\hat{D}_-(\alpha_-)\hat{S}_{+-}(\zeta)|0\rangle|0\rangle$, where $\hat{D}_\pm(\alpha_\pm)=\exp(\alpha_\pm \hat{a}_\pm^\dagger - \alpha_\pm^* \hat{a}_\pm)$ is the displacement operator for the two phonon modes described by annihilation operators $\hat{a}_+$ and $\hat{a}_-$, $\hat{S}_{+-}(\zeta) = \exp(\zeta^* \hat{a}_+ \hat{a}_- - \zeta \hat{a}_+^\dagger \hat{a}_-^\dagger)$ is the unitary two-phonon squeeze operator with the complex squeeze parameter $\zeta = s\exp(i\theta)$, and $|0\rangle|0\rangle$ is the two-phonon vacuum state [6] as schematically shown in the inset of Fig.4. A key feature of squeezing is that it results in a redistribution of uncertainty between quadratures. The squeezing induces correlations between orthogonal quadratures of two separate phonon modes by mixing the annihilation operators $\hat{a}_{\pm}$ of one phonon mode with the creation operators $\hat{a}_{\mp}^{\dagger}$ of another one. The two-mode squeezed state is analogous to a thermal state as both states satisfy the same relations between the expected number of phonons and the probability of finding *n* phonons in *any* of the modes This means that disregarding one of the modes or looking at the reduced density matrix of the composite system in one subspace we will see the phonon ensemble with thermal statistics.

### 3.2 Coherent control experiments

A well defined phase of coherent lattice excitations can be used directly for a variety of manipulations aiming at the preparation of a preferred quantum state. This can be achieved by coherent control technique the main idea of which is to generate a superposition of lattice states with well defined relative phases. By varying the phases, it is possible to bring the superposition either to a destructive or a constructive interference and thus to control the final state reached after the excitation. Such manipulation can be accomplished by two-pump, one-probe technique. However, since we deal with composite phonons having two phases: one for the composite phonon itself and another for its constituent part, our coherent control experiment is an acoustical analogue of two-photon interference for parametrically downconverted photons [7, 32]. Indeed, instead of overlapping pair-correlated photons, we superimpose two ensembles of biphonons created at different times as schematically shown in Fig.4, where the coherent control experiment is depicted in simple conceptual terms. By the coherent control, we can steer the lattice either into biphonon state where both modes are simultaneously excited, or into two-phonon vacuum in which they are empty. The two-phonon vacuum corresponds then to the situation when all atoms are almost at equilibrium position (the distribution of atomic displacements is narrow), while their kinetic energy is maximal due to the broad momentum distribution. In contrast, in two-phonon state the atoms are at rest (the velocity distribution is narrow), but the displacement distribution is wide.

As the first approximation, the whole coherent control process can be described as the sum of two biphonon ensembles whose motion is initiated at different times and that now interfere. The relative timing of the two pump pulses determines whether the oscillations resulting from the biphonon state, in which the constituent phonons are in a superposition state, add constructively or destructively. Figure 5(a) displays the modification of biphonon oscillations when interpulse separation is varied. As can be seen from this figure, the coherent biphonon amplitude in ZnTe changes systematically in such a way that that for a fixed interpulse separation, the oscillatory signal can be either significantly enhanced or almost suppressed. As shown in Fig.5 (b), the resulting amplitude of biphonon oscillations is harmonically modulated depending on the interpulse separation in such a way that its modulation period coincides with the biphonon period. In stark contrast to the biphonon amplitude that behaves classically, the observed modulation of lifetime is difficult to explain within a classical interference scheme. The change of lifetime observed in our experiments can be either correlated or anti-correlated to that of amplitude, and the decrease in lifetime, with the ratio reaching $1.5 \pm 0.2 \approx \sqrt{2}$, occurs over a shorter time scale as compared to its increase, which takes place more smoothly. This modulation of lifetime, shown in Fig.5(b), unequivocally suggests that we are dealing with quantum interference (that is the interference of superpositions since each constituent of the biphonon state is a superposition of its vacuum and first excited state). Indeed, for classical interference, the change of coherence lifetime is impossible: classical coherent states are always transformed into different coherent states. The biphonon amplitude scales with the real part of squeeze parameter [12, 13], whereas the oscillation lifetime measures how long the constituents of biphonon state are correlated or, may be entangled: the longer the time, the stronger the correlation. Here we must emphasize that despite the fact that in our experiments the squeezing factor (proportional to biphonon amplitude) is one order of magnitude larger than for $KTaO_3$ [12], we still cannot *experimentally* prove that we have achieved vacuum squeezing, even though we do believe that the effects observed are due to quantum, not thermal (classical) fluctuations. The problem here is that there is no reliable reference relative to which one can estimate the size of fluctuations, for more details, see [10, 11]. However, for the measurements on ultrafast scale the criterion for the system to behave quantum mechanically [33] is $kT \leq \frac{\hbar \omega}{2} \frac{\tau^*}{\tau}$, where T is the temperature, $\omega$ and $\tau^*$ are the characteristic frequency and relaxation time of the system, respectively, while $\tau$ is the measurement time. Given the characteristic relaxation time for acoustical phonons of 10 ps, and the measurement time of 40 fs, the criterion is satisfied already at room temperature. Thus, the behavior of lattice on such short time is primarily dominated by quantum fluctuations.

Let us consider the coherent control of biphonons in more detail. First pump pulse creates a biphonon state that freely evolves in time until at time delay $\Delta t$ it is overlapped with second biphonon state. The phase difference for the two composite states is defined to be the interpulse separation $\Delta t$. The internal phase $\phi_i$ for each biphonon state is defined to be the phase difference of the constituent states and it has no direct analogue for classical interference. When the interpulse separation is a multiple of biphonon period, the resulting internal phase $\Delta\phi = \phi_1 - \phi_2$ equals zero. This corresponds to the situation when two biphonon ensembles are superimposed with the same orientations of uncertainty ellipse. However, for $\Delta t \neq nT$ where $n$ is the integer, the orientations are different, thus the internal phase $\Delta\phi = \Delta t - nT$ can be either positive or negative depending on whether $n$ is even or odd integer. Thus, the sign of internal phase $\text{sgn}(\Delta\phi)$ is responsible for an abrupt change of lifetime near $\Delta t/T = 2.5$. In total, on the left of $\Delta t/T = 2.5$ the lattice excitation behaves as a collection of *independent* anharmonic oscillators for each of which the decay is correlated with its biphonon amplitude. At the same time, on the right side of this border line the lattice excitation resembles a set of *coupled* harmonic oscillators with the decay controlled by their coupling (dispersion) inversely proportional to the biphonon amplitude. Were one able to measure the statistics of the constituent phonons it would exhibit antibunching on the left and bunching on the right of the border line $\Delta t/T = 2.5$, both evolving into the Poissonian statistics at $\Delta t/T = 2$ or $\Delta t/T = 3$. Thus, on the left of the border line, where the energy is predominantly localized on the lattice atoms, the system behavior is particle-like, whereas on the right, when the energy is delocalized, it demonstrates wave-like behavior. Moreover, the entanglement on the left is primarily controlled by amplitude fluctuations, while on the right by phase fluctuations. Only at the border line, where the phase and amplitude noises act together we observe the sudden change in lifetime caused by the fact that the fluctuations affect localized and distributed coherences in very different ways [34].

Naively, one might think that a larger squeezing always results in a stronger entanglement. However, as follows from Fig. 5(b), the maximal lifetimes come about near a *minimum* of the resulting biphonon amplitude, where the squeezing has to be minimal (the squeezing factor $|s| \approx 1$ means that the fluctuations are the same for both quadratures). The peculiar lifetime dependence reflecting the particle-wave duality together with the regular amplitude dependence illustrate that the relationship between the phonon squeezing and entanglement may be more complicated because these effects are of different orders. The squeezing is the second order effect relying on the amplitude-amplitude correlations primarily controlled by one-phonon interference. The entanglement, in contrast, belongs to a class of the forth order effects governed by intensity-intensity correlations. This is the same class as bunching/anibunching phenomena dictated by field statistics [7, 8, 32]. Here it is appropriate to note that any generalized coherent state realized for parametric excitation of a quantum system can be either squeezed or correlated. However, as shown in [35], these states can be physically and mathematically different as appear, for instance, in spontaneou**s** emission**.**

Dissimilarity in the behavior of biphonon amplitude and lifetime can be better understood by noting that the squeezing affects [6] only the diagonal phonon-number expectation values for each individual phonon mode $<\hat{a}^\dagger_\pm \hat{a}_\pm> = |\alpha_\pm|^2 + \sinh^2 s$ and the off-diagonal intermode expectation values $<\hat{a}_+ \hat{a}_-> = <\hat{a}_- \hat{a}_+> = \alpha_+ \alpha_- - \exp(i\theta)\sinh(s)\cosh(s)$. Therefore, for a larger squeezing the population of each phonon mode can only increase, whereas the mode entanglement can either increase or decrease. Loosely speaking, two-phonon interference modulates both the phase $\theta$ and the amplitude $s$ of complex squeeze parameter, whereas one-phonon interference is only responsible for its amplitude modulation. Thus, we see that similar to the case of photons [34], one- and two-phonon interference is controlled by coherence and correlation/entanglement, respectively. One-phonon interference tends to localize the modes in the phase space, whereas two-phonon interference affects their overlap. Depending on interpulse separation we can demonstrate the change from one limit to the other since two-phonon interference oscillates with twice the frequency as one-phonon interference. As distribution in the phase space for biphonon state becomes broader, one-phonon coherence is decreased while correlation is enhanced, so that the strength of one-phonon interference decreases while that of two-phonon interference increases. Because of the opposite dependence of coherence and correlation /entanglement on separability, the distribution size plays opposite roles in determining the strength of one- and two-phonon interference. A large size corresponds to highly correlated/entangled phonon pairs with low degree of coherence, whereas smaller size results in highly coherent phonons that are poorly correlated/entangled. Concluding, it could not be helped noticing that we currently cannot quantify either entanglement or squeezing. It is just taken for granted that the first property is proportional to lifetime, whereas the second - to amplitude of the biphonon oscillations as suggested by earlier research studies [12,13]. Thus, we have left open the question of whether or not our measurements dealt with real *vacuum* squeezing and true *quantum* correlations.

## 4.    Conclusions

To summarize, we have demonstrated that ultrafast laser pulses provide a flexible and powerful tool not only to create and observe but also to control phonon squeezing and correlation. Besides representing a landmark observation in quantum acoustics and being the first experimental demonstration of the complementarity between one- and two-phonon interference in the time domain, our study opens new possibilities for the exploration and exploitation of phonon squeezing and correlation/entanglement effects. Owing to their remarkable properties, the correlated and squeezed phonons can be used in a broad variety of experiments that extend from the fundamental to the applied.

**Acknowledgements**

The authors appreciate financial support from the Ministry of Education, Culture, Sports, Science and Technology of Japan, and are thankful to Hiroshi Takahashi for the help in designing the experimental setup.

FIGURE CAPTIONS

Fig.1. (colour on-line) Schematic of the experimental setup. $M_i$ denote the mirrors, BS - the beam splitter, P - the polarizer, L - the lens, $PD_i$ - the photodetectors. The delay line is made of a pair of mirrors, M1 and M2, mounted on a shaker oscillating at 20 THz. A modified Mach-Zehnder interferometer is used to make coherent control experiments.

Fig. 2. (colour on-line) Single pump excitation. (a) - Typical transient transmission change of (001) ZnTe consisting of coherent biphonon oscillations. The oscillation lifetime is 1.4 ps, as can be deduced from either the autocorrelation function depicted in the panel (b) or a fit in real time. The oscillation frequency of 3.67 THz is derived from Fourier transformed autocorrelation function shown in the panel (c).

Fig.3. (colour on-line) Sketch of the excitation process. The unitary cell of a non-centrosymmetric crystal containing two atoms is shown. The averaged interparticle separation between the atoms is $r$, whereas the instantaneous displacement from the equilibrium position is $u$. The filled circles represent the uncertainty equal to the variance $\langle u^2 \rangle$. The forces acting on the atoms in each TA(X) phonon mode are schematically presented by coloured arrows.

Fig. 4. (colour on-line) Schematic of the coherent control experiment. When the interpulse separation is a multiple of biphonon period, the resulting internal phase $\Delta\phi = \phi_1 - \phi_2 \equiv 0$. However, for $\Delta t \neq nT$ the internal phase $\Delta\phi = \Delta t - nT$ can be either positive or negative depending on whether $n$ is even or odd integer. In the insert, the two-phonon squeezed vacuum is represented by an uncertainty ellipse, illustrating how the squeezing amplitude $s$ and phase $\theta$ determine the eccentricity and orientation of the ellipse. The squeezing takes place not in the individual modes but in the total quadrature components defined here as the sum $Y_1$ and difference $Y_2$ of individual quadratures with the squeezing conditions $\cos\theta > \tanh s$ and $\cos\theta < -\tanh s$, respectively. The dotted circle represents the vacuum fluctuations.

Fig. 5. (colour on-line) Experimental results of coherent control. (a) – Double-pulse excitation data demonstrating enhancement and suppression of biphonon oscillations. Zero delay time corresponds to the arrival of second pump pulse. The curves are offset along the vertical axis for clarity. (b) – Amplitude (closed symbols) and lifetime (open symbols) of biphonon oscillations as a function of interpulse separation. The amplitude and lifetime are obtained from fitting the time-domain data and extrapolating the oscillations to the zero delay time.

# References

1. L. Dhar, J.A. Rogers, K.A. Nelson, Chem. Rev. 94 (1994) 157-193.

2. R. Merlin, Solid State Commun. 102 (1997) 207-220.

3. T. Dekorsy, G.C. Cho, H. Kurz, in: M. Cardona, G. Güntherodt (Eds), Light Scattering in Solids VIII, Springer, Berlin, 2000, pp. 169-208.

4. O.V. Misochko, Zh. Eksp. Teor. Fiz. **119** (2001) 285-300 [JETP 92 (2001) 246-259].

5. K. Ishioka, O.V. Misochko, in: K. Yamanouchi, A. Giullietti, K. Ledingham (Eds), Progress in Ultrafast Intense Laser Science **V**, Springer Series in Chemical Physics, Berlin, 2010, pp. 23-64.

6. R. Loudon, P.L. Knight, Journal of Modern Optics 34 (1987) 709-759.

7. D.N. Klyshko, Photons and Nonlinear Optics, Gordon and Breach, New York, (1988).

8. C.C. Gerry, P.L. Knight, Introductory Quantum Optics, Cambridge University Press, Cambridge, U.K., 2005.

9. H. Hu, F. Nori, Phys. Rev. Lett. 76 (1996) 2294-2297.

10. H. Hu, F. Nori, Physica B 263-264, (1999) 16-29.

11. S. Sauer, J.M. Daniels, D.E. Reiter, T. Kuhn, A. Vagov, V.M. Axt, Phys. Rev. Lett. 105 (2010) 157401.

12. G. Garrett, A.A. Rojo, A.K. Sood, J.F. Whitaker, R. Merlin, Science 275 (1997) 1638–1640.

13. G.A. Garrett, J.F. Whitaker, A.K. Sood, R. Merlin, Opt. Express 1 (1997) 385-389.

14. O.V. Misochko, K. Sakai, S. Nakashima, Phys. Rev. B 61 (2000) 11225-11228.

15. A. Bartels, T. Dekorsy, H. Kurz, Phys. Rev. Lett. 84 (2000) 2981–2984.

16. O.V. Misochko, Phys. Lett. A 269 (2000) 97–102.

17. O.V. Misochko, K. Kisoda, K. Sakai, S. Nakashima, Appl. Phys. Lett. 76 (2000) 961-963.

18. S.L. Johnson, P. Beaud, E. Vorobeva, C.J. Milne, É.D. Murray, S. Fahy, G. Ingold, Phys. Rev. Lett. 102 (2009) 175503.

19. E.S. Zijlstra, L.E. Díaz-Sánchez, M.E. Garcia, Phys. Rev. Lett. **104** (2010) 029601.

20. A. Hussain, S. R. Andrews, Phys. Rev. B 81 (2010) 224304.

21. A. Einstein, B. Podolsky, N. Rosen, Phys. Rev. 47 (1935) 777-780.

22. E. Schrödinger, Naturwissenschaften 23 (1935) 807-812, 823-829, 844-849. A translation of these papers appears in 1983, *Quantum Theory and Measurement*, J.A. Wheeller, W.H. Zurek (Eds), Princeton: Princeton Univ.

23. V.M. Agranovich, I.I. Lalov, Uspekhi Fiz.Nauk 146 (1985) 267-302 [Sov.Phys.Uspekhi 28 (1985) 484].

24. J. Hu, O.V. Misochko, N. Takei, K. Ohmori, K. G. Nakamura, preprint at http://arxiv.org/abs/ 1011.6115 (2010).

25. R. Scholz, T. Pfeifer, H. Kurz, Phys. Rev. B 47 (1993) 16229-16236.

26. A. V. Kuznetsov, C. J. Stanton, Phys. Rev. Lett. 73 (1994) 3243-3246.

27. M.H. Cohen, J. Ruvalds, Phys. Rev. Lett. 23 (1969) 1378-1381.

28. J.C. Kimball, C.Y. Fong, Y.R. Shen, Phys. Rev. B 23, 4946-4959 (1981).

29. R.M. Shafroth, Phys. Rev. 96 (1954) 1149.

30. A. Luis, Phys. Rev. A 64 (2001) 054102.

31. Z. Ivica, G.P. Tsironisb, Physica D 216 (2006) 200–206.

32. L. Mandel, Reviews of Modern Physics 71 (1999) S274-282.

33. V.B. Braginsky, F.Ya. Khalili, Quantum measurement, Cambridge University Press, Cambridge, U.K. (1992).

34. A.F. Abouraddy, M.B. Nasr, B.E.A. Saleh, A.V. Sergienko, M.C. Teich, Phys. Rev. A 63 (2001) 063803.

35. V.V. Dodonov, E.V. Kurmyshev, V.I. Man'ko, Phys. Lett. A 79 (1980) 150–152.

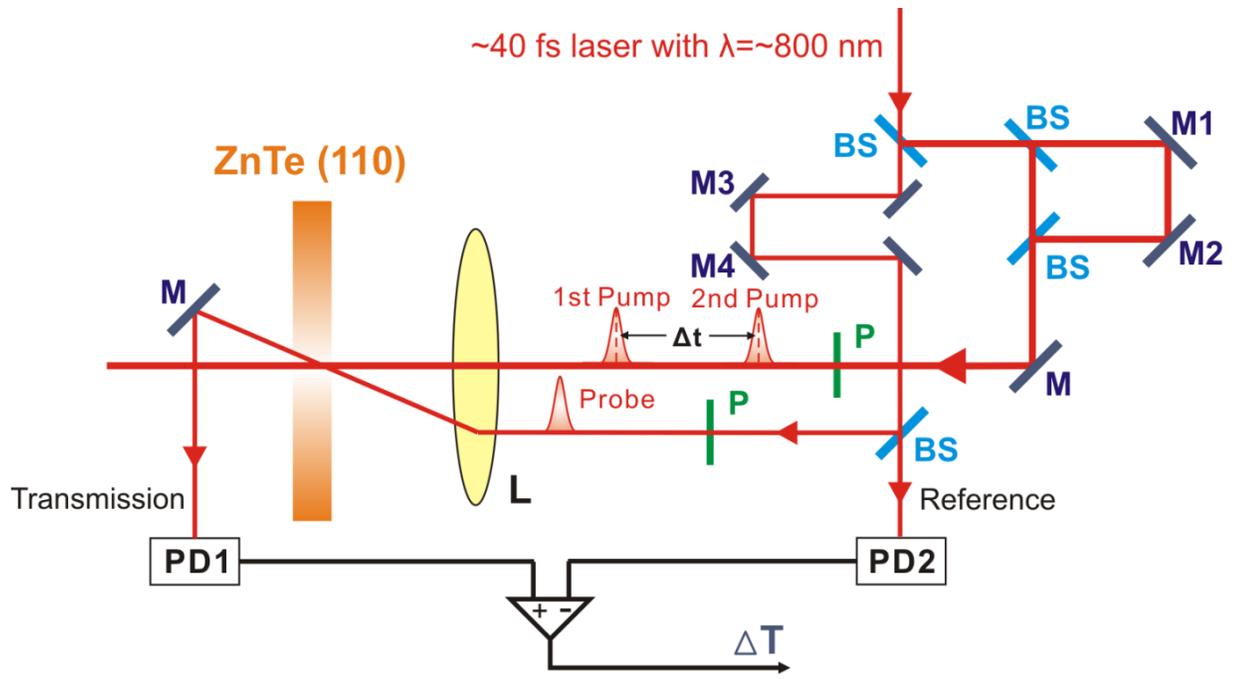

Figure 1.

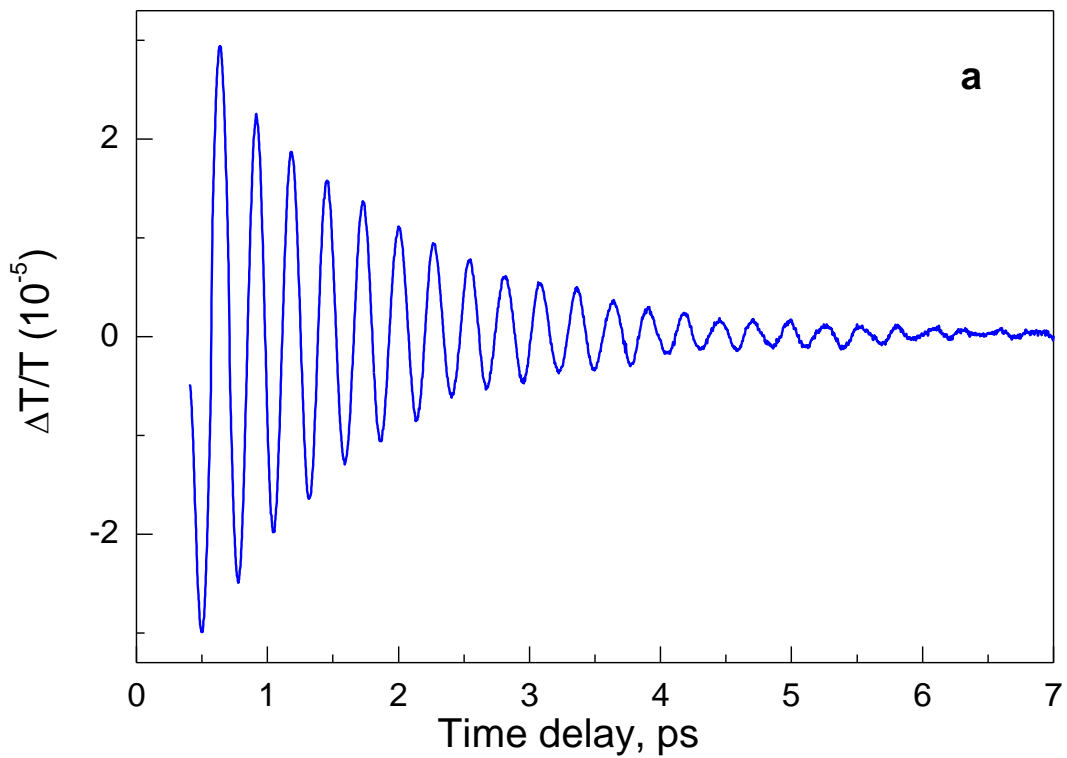
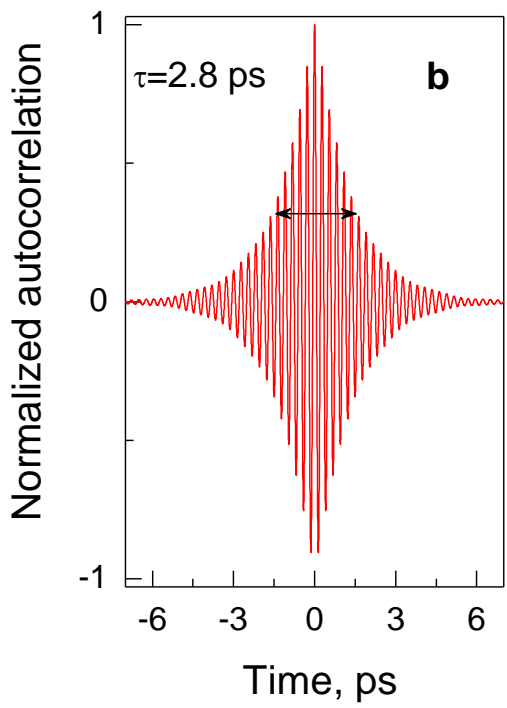
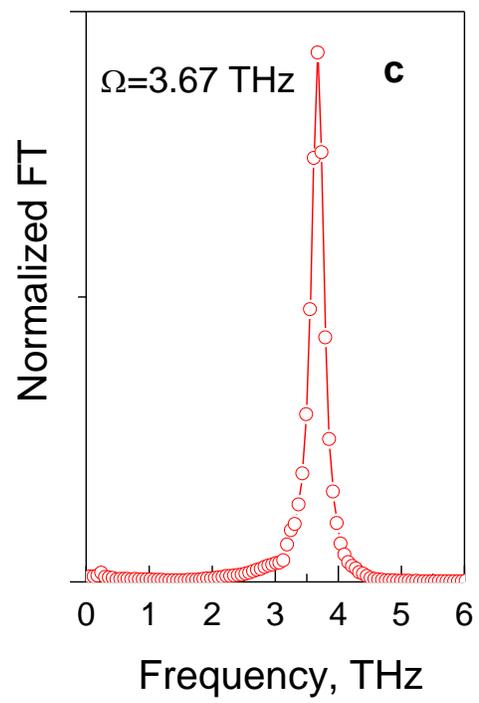

Figure 2.

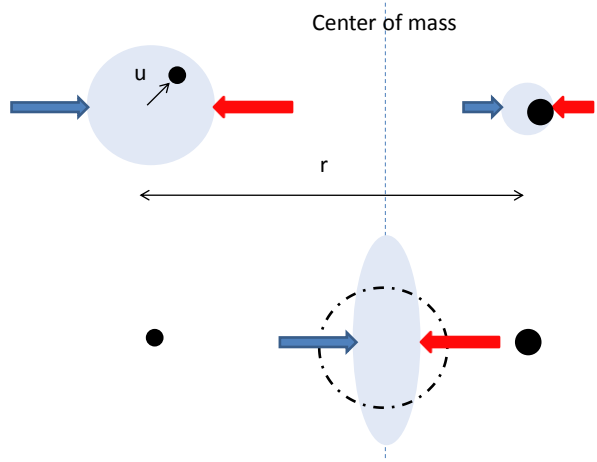

Figure 3.

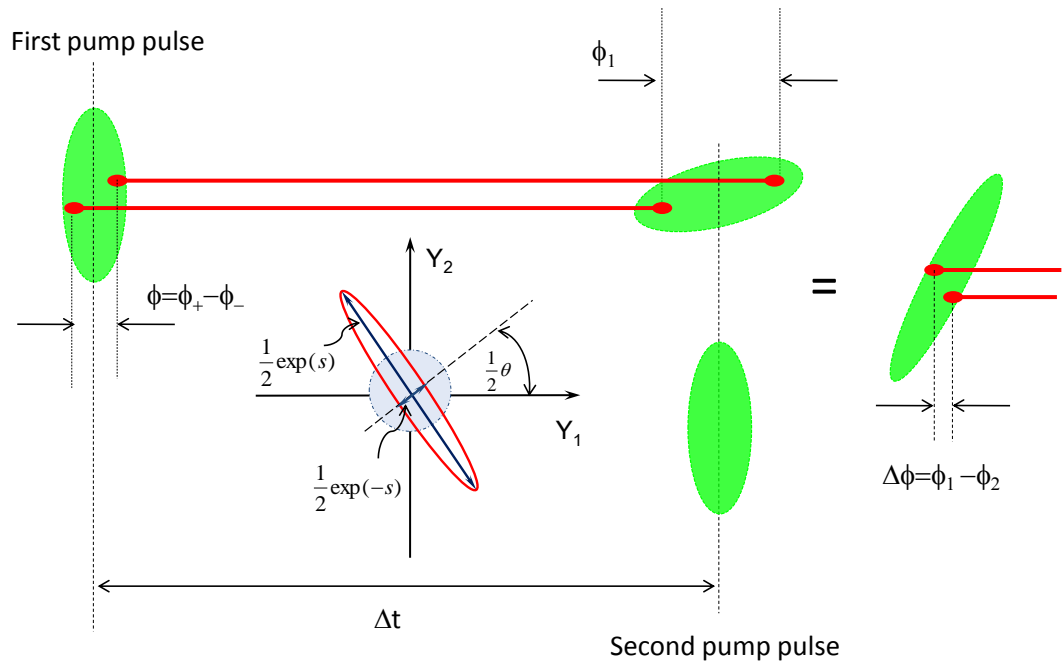

Figure 4.

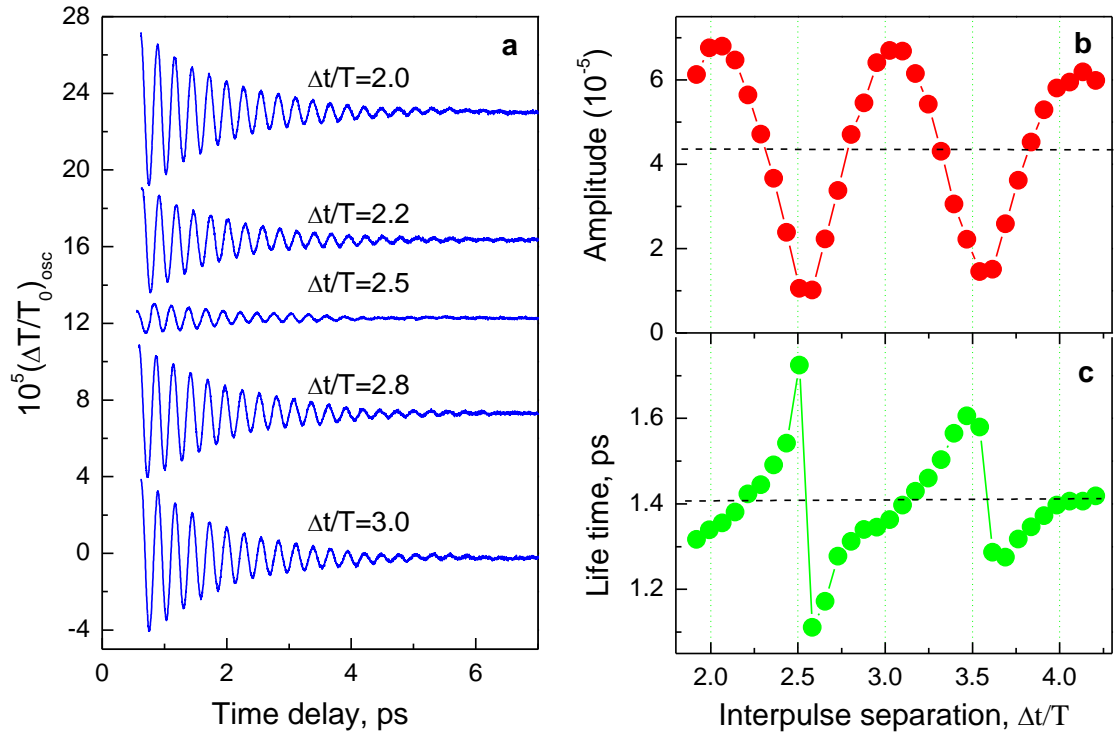

Figure 5.